\documentclass[sn-mathphys,Numbered]{sn-jnl}

\usepackage{graphicx}%
\usepackage{multirow}%
\usepackage{amsmath,amssymb,amsfonts}%
\usepackage{amsthm}%
\usepackage{mathrsfs}%
\usepackage[title]{appendix}%
\usepackage{xcolor}%
\usepackage{textcomp}%
\usepackage{manyfoot}%
\usepackage{booktabs}%
\usepackage{algorithm}%
\usepackage{algorithmicx}%
\usepackage{algpseudocode}%
\usepackage{listings}%
\usepackage{soul} 
\usepackage{array}
\usepackage{comment}

\theoremstyle{thmstyleone}%
\theoremstyle{thmstyletwo}%

\theoremstyle{thmstylethree}%

\begin{document}

\title[Article Title]{How quantum computing can enhance biomarker discovery}

\author*[1,2]{\fnm{Frederik F.} \sur{Flöther}}\email{frederik.floether ``at'' quantumbasel.com}

\author[3]{\fnm{Daniel} \sur{Blankenberg}}

\author[4,5]{\fnm{Maria} \sur{Demidik}}

\author[4,5]{\fnm{Karl} \sur{Jansen}}

\author[6]{\fnm{Raga} \sur{Krishnakumar}}

\author[1,2]{\fnm{Rajiv} \sur{Krishnakumar}}

\author[7]{\fnm{Numan} \sur{Laanait}}

\author[8]{\fnm{Laxmi} \sur{Parida}}

\author[3]{\fnm{Carl} \sur{Saab}}

\author[8]{\fnm{Filippo} \sur{Utro}}

\affil[1]{\orgname{QuantumBasel}, \orgaddress{\street{Schorenweg 44b}, \city{Arlesheim}, \postcode{4144}, \country{Switzerland}}}

\affil[2]{\orgname{Center for Quantum Computing and Quantum Coherence (QC2), University of Basel}, \orgaddress{\street{Petersplatz 1}, \city{Basel}, \postcode{4001}, \country{Switzerland}}}

\affil[3]{\orgname{Cleveland Clinic}, \orgaddress{\street{Euclid Ave 9500}, \city{Cleveland}, \postcode{44195}, \country{USA}}}

\affil[4]{\orgname{Deutsches Elektronen-Synchrotron DESY}, \orgaddress{\street{Platanenallee 6}, \city{Zeuthen}, \postcode{15738}, \country{Germany}}}

\affil[5]{\orgname{Computation-Based Science and Technology Research Center, The Cyprus Institute}, \orgaddress{\street{Kavafi Street 20}, \city{Nicosia}, \postcode{2121}, \country{Cyprus}}}

\affil[6]{\orgname{Sandia National Laboratories}, \orgaddress{\street{East Avenue 7011}, \city{Livermore}, \postcode{94550}, \country{USA}}}

\affil[7]{\orgname{Carelon, Elevance Health}, \orgaddress{\street{Virginia Avenue 220}, \city{Indianapolis}, \postcode{46204}, \country{USA}}}

\affil[8]{\orgname{IBM Thomas J. Watson Research Center}, \orgaddress{\street{Kitchawan Road 1101}, \city{Yorktown Heights}, \postcode{10598}, \country{USA}}}



\maketitle

\begin{abstract}

Biomarkers play a central role in medicine's gradual progress towards proactive, personalized precision diagnostics and interventions. However, finding biomarkers that provide very early indicators of a change in health status, for example for multi-factorial diseases, has been challenging. Discovery of such biomarkers stands to benefit significantly from advanced information processing and means to detect complex correlations, which quantum computing offers. In this perspective paper, quantum algorithms, particularly in machine learning, are mapped to key applications in biomarker discovery. The opportunities and challenges associated with the algorithms and applications are discussed. The analysis is structured according to different data types --- multi-dimensional, time series, and erroneous data --- and covers key data modalities in healthcare --- electronic health records (EHRs), omics, and medical images. An outlook is provided concerning open research challenges.

\end{abstract}

\section{Introduction}
\label{sec:intro}

A biomarker is a defined characteristic that is measured as an indicator of normal or pathological biological processes, or of response to an exposure or intervention \cite{working2001biomarkers}. The study of biomarkers arguably dates back to at least the beginning of the last century with Karl Landsteiner's discovery of the ABO blood types in 1901 (recognized with the 1930 Nobel Prize in Physiology or Medicine). Biomarkers consist of both biological molecules \cite{henry2012cancer} and other medical signs that indicate diseases \cite{strimbu2010biomarkers, aronson2017biomarkers}, in contrast to symptoms as well as more general medical endpoints. Biomarkers can be based on histological, molecular, physiological, and radiographic characteristics. They may also take the form of aggregate high-dimensional features.

Common modalities of biomarkers include biomolecules, fields in electronic health records (EHRs), laboratory values, medical images, omics information, social determinants, and wearable data. Moreover, there are several application categories of biomarkers, including diagnostic, monitoring, prognostic, predictive, response, safety, and susceptibility/risk \cite{group_glossary_2021}. Hence, biomarkers are context-dependent and use case-specific in healthcare and clinical trial settings.

For example, clinical trials in neurology and psychiatry are time-intensive, expensive, and subjective, and have lower success rates than other indications \cite{thomas2016clinical}. Within neuro-psychiatry, pain represents a large unmet need with respect to the lack of reliable biomarkers. Hence, quantitative and physiological biomarkers are urgently needed for a comprehensive and reliable diagnosis and assessment of pain in healthcare and clinical trials \cite{davis2020discovery}. This is illustrative of the impact better biomarkers can have across medicine.

While the beginnings of quantum mechanics date back 100 years, quantum computing is a new form of information processing, including novel hardware and software, that leverages quantum mechanical effects such as quantum entanglement, interference, and superposition. It is the only known form of computing that can provide, at least for certain problems, exponential speedups compared with classical approaches; for other problems, there may only be polynomial speedups or none at all, however \cite{horowitz2019quantum}. Hundreds of use cases are meanwhile being explored across industry and academia \cite{langione2023quantum}. Quantum computing is not just a faster way of addressing problems; it represents an entirely different way to find solutions. The benefits also go beyond speed; depending on the application and the use case, the main benefit could be accuracy \cite{peral2024systematic}, energy efficiency \cite{jaschke2023quantum}, or the ability to handle difficult datasets (for instance, high-dimensional or highly noisy ones) \cite{peters2021machine, marshall2023high}.

The focus of this review is on (gate-based) quantum computing and its potential for computational problems in biomarker discovery. Nevertheless, it is important to note that there are further quantum technologies with medical applications and biomarker relevance, in particular quantum annealing \cite{felefly2023explainable} and quantum sensing \cite{mauranyapin2022quantum, flother2023can, shams2023quantum}. In the rest of the paper, quantum computing and relevant quantum algorithms are first summarized in Section \ref{sec:qc} before the application of quantum algorithms to biomarker discovery problems is described in Section \ref{sec:qcbio}. Open research challenges are discussed in Section \ref{sec:open} and conclusions are drawn in Section \ref{sec:conc}.

\section{Quantum computing}
\label{sec:qc}

Quantum computing is a relatively nascent field of research undergoing rapid and substantial developments. By leveraging the fundamental properties of quantum mechanics, namely superposition and entanglement, quantum computing presents exciting opportunities to solve problems more efficiently than classical computing can. However, currently available quantum computers are prone to errors and face limitations in both the number of quantum bits (qubits) and the connectivity between them.  Overcoming these constraints is essential to implementing and correctly executing any algorithm. These constraints also introduce challenges in developing algorithms to achieve a quantum advantage with available quantum devices, commonly referred to as utility-scale or near-term quantum computers. Therefore, significant efforts concentrate on utilizing near-term quantum computers, while research continues on developing fault-tolerant quantum hardware that can operate reliably even in the presence of noise.

In quantum computing, algorithms take advantage of Hilbert space. This is a high-dimensional vector space that represents the state of the qubits and is naturally suited to computing outcome probabilities (via an inner product). If the input of a quantum algorithm is classical data that is not represented by quantum states, data embedding is required in order to perform operations on a quantum computer. While there are various ways to embed classical data into quantum states, the number of available qubits restricts the feature space size of the data we can tackle with near-term devices. For instance, angle encoding~\cite{schuld2018supervised} or instantaneous quantum polynomial encoding~\cite{havlicek_supervised_2019} require a linear scaling of the number of qubits with respect to the number of features. Alternatively, amplitude encoding provides a logarithmic scaling in the number of qubits~\cite{schuld2018supervised}. However, without ancillary qubits, amplitude encoding results in deep circuits~\cite{nakaji_amplitude_2022}. In practice, when the feature space exceeds the capacity of available quantum hardware, classical dimensionality reduction algorithms are applied as a pre-processing step in quantum pipelines. Still, with ongoing advancements in quantum computing, biomarker discovery and other quantum applications may eventually be performed without the need for significant (classical) pre-processing.
 
Quantum circuits are fundamental to the architecture of quantum computation, providing the framework through which quantum algorithms are designed and executed. At a high level, a quantum circuit is a sequence of quantum gates applied to qubits. Each gate performs a specific operation on one or more qubits. Fault-tolerant quantum algorithms typically require running deep quantum circuits on tens of thousands to millions of qubits, making them unsuitable for near-term devices. On the other hand, variational quantum algorithms (VQAs)~\cite{cerezo_variational_2021} utilize shallower circuits and can produce meaningful results. This characteristic has made VQAs highly popular for near-term quantum computing applications. VQAs are hybrid quantum-classical algorithms that involve an iterative loop between quantum and classical hardware. First, quantum circuits that have parametrized gates, called the ans\"atze or parametrized quantum circuits (PQCs), are employed to obtain samples from a quantum computer. Then, a classical computer optimizes the parameters of the quantum circuit with respect to the loss function of the problem. Most quantum machine learning models belong in the class of VQAs and can be evaluated on quantum computers. Alternatively, quantum computers could be leveraged as sub-routines within classical algorithms. One example is the thermal state preparation during Boltzmann machine training. However, the performance of VQAs is highly sensitive to hardware noise, necessitating the use of error mitigation techniques to partially counteract noise-induced errors~\cite{cai_quantum_2023}.

The execution of quantum circuits is performed on quantum hardware through quantum processing units (QPUs), in analogy to central (CPUs) and graphics processing units (GPUs). Another element is the quantum random access memory (QRAM), which is a quantum memory that stores a quantum state for later retrieval. Unlike QPUs, which are readily available via cloud computing platforms, hardware incorporating QRAM devices has not yet been demonstrated.

Quantum computing is naturally suited for quantum data (which describes quantum states) \cite{luongo2020quantum}. However, it also holds great promise in solving problems using classical data too. In particular, there has been much progress in the last few years in understanding how quantum computers can solve computational problems that can be formulated as cryptography, machine learning (ML), optimization, or simulation (of nature) problems \cite{2021}. In this paper, we focus on analyzing classical data with the help of quantum algorithms, including multi-dimensional, time series, and erroneous data, which is discussed in detail in Section~\ref{sec:qcbio}.

\subsection{Quantum machine learning}

Quantum machine learning (QML) integrates principles from quantum computing and machine learning, promising potential advances in learning from data. A recent benchmarking study~\cite{bowles2024better} provides a comprehensive evaluation of popular QML models on binary classification problems. The study examines QML architectures compatible with near-term quantum devices such as quantum neural networks (QNNs), including quantum convolutional neural networks (QCNNs) and quantum kernel methods (QKMs).

The majority of QNNs are based on PQCs. The parameters of PQCs are optimized following the VQA paradigm. A typical QNN begins with data embedding, followed by a series of gates. One of the most general QNN models for classification is the variational quantum classifier (VQC), which is essentially a PQC. Another example of a VQC is the circuit-centric quantum classifier~\cite{schuld_circuit-centric_2020}, one of the earliest proposed generic QNNs. The data re-uploading classifier is a notable QNN architecture that was shown to have universal approximation properties~\cite{perez-salinas_data_reup_2020}. The data re-uploading classifier features a processing layer composed of data encoding and unitary operations with trainable parameters. This layer is repeated a number of times, each layer having independent parameters.

Recently, significant attention has been given to leveraging quantum computers for evaluating kernel functions in kernel methods. Quantum kernels, based on feature maps (data encodings), could be used with classical maximum margin classifiers. To evaluate a quantum kernel function, a circuit is designed by first applying a feature map and then its Hermitian conjugate. A prominent example of data-dependent quantum advantage in QML models is demonstrated through QKMs. The authors of the manuscript~\cite{huang_power_2021} established a relation between the potential advantage in predictive accuracy of a model for a given dataset and a quantum kernel. In addition, a projected quantum kernel classifier was introduced.

Although the predictive advantage of QML models for practically relevant problems remains an open question and necessitates data embeddings that are difficult to simulate classically, quantum models are expected to show better generalization than classical models, leveraging fewer data points~\cite{caro2022generalization}. This is particularly intriguing in fields where constructing sufficiently large datasets to ensure satisfactory performance for ML models is challenging, as is the case in many healthcare settings.

The findings in~\cite{bowles2024better} highlight the significance of experimental design for the performance of quantum models. Problem-agnostic implementations of QML models are unlikely to outperform their classical counterparts, highlighting the need for tailored approaches. Specifically, the embedding of classical data into Hilbert space and subsequent operations should be tailored to an application and available hardware. Additionally, for QML models based on VQAs, the barren plateau (BP) phenomenon should be addressed. A model experiences a BP when its loss landscape flattens, and the variance of parameters' gradients decays exponentially as the system size increases, making gradient-based optimization challenging. Avoiding BPs in QML models commonly necessitates reducing the expressivity of the circuits, which also tends to make them more classically simulable (decreasing the potential of significant quantum advantages)~\cite{cerezo_does_2024, larocca_review_2024}. These insights underscore the importance of use case-specific algorithm design for the potential benefits of quantum computing. 

Despite the challenges in algorithm design, a variety of (machine) learning tasks are being explored with quantum algorithms. An overview of learning tasks, use case examples, and corresponding quantum algorithms (many of the algorithms may also be relevant to other learning tasks and use cases) is given in Table \ref{table:learn} and will now be further discussed.

\begin{table}
    \centering
    \begin{tabular}{ m{3.7cm}  m{4cm} m{4cm} }
       Learning task  & Use cases & Quantum algorithms \\
    \hline
    \hline
         Dimensionality reduction & Covariates for biomarker trait regression models & QPCA~\cite{lloyd_qPCA, covariance_qPCA}, QLDA~\cite{yu_qLDA_2023}, QSFA~\cite{kerenidis_classification_qSFA_2020}, QIsomap~\cite{feng_qIsomap_2024} \\
         Classification & Early disease stage prediction & QNNs~\cite{qcnn, perez-salinas_data_reup_2020, schuld_circuit-centric_2020, ray2023hybrid}, QKMs~\cite{havlicek_supervised_2019, huang_power_2021} \\ 
         Regression & Disease risk prediction & Quantum linear regression~\cite{wiebe_quantum_fitting_2012, wang_quantum_regression_2017, zhao_quantum-assisted_gpr_2019, date_adiabatic_regression_2021} \\
         Clustering & Multi-omics data analysis & QK-means~\cite{khan_qk-means_2019, arthur_balanced_k_means_2021}, quantum spectral clustering~\cite{kerenidis_quantum_spectral_2021} \\
         Generative learning & Synthetic medical images & QNNs~\cite{tian2023recent}, QBMs~\cite{amin_quantum_2018, tuysuz_learning_2024}, QGANs \cite{baglio2024data} \\ 
         Time series forecasting & Longitudinal patient studies, drug resistance mechanisms & QRC \cite{suzuki2022natural} \\
         Natural language processing & Clinical notes processing & QLSTM~\cite{di2022dawn} \\ 
         Optimization & Treatment optimization & QAOA~\cite{choi2019tutorial}, QPSO~\cite{flori2022quantum}, VQE~\cite{tilly2022variational} \\
         \hline
    \end{tabular}
    \caption{Overview of machine learning and related tasks and examples of use cases and quantum algorithms.}
    \label{table:learn}
\end{table}


Consider the example of dimensionality reduction, a key technique in order to compress large and complex datasets while preserving key information. As quantum computing rapidly develops, quantum algorithms for reducing the feature space size are evolving alongside it. For instance, many classical dimensionality reduction algorithms have a quantum adaptation, such as quantum principal component analysis (QPCA)~\cite{lloyd_qPCA}, quantum linear discriminant analysis (QLDA)~\cite{yu_qLDA_2023}, quantum slow feature analysis (QSFA)~\cite{kerenidis_classification_qSFA_2020}, and quantum isomap (QIsomap)~\cite{feng_qIsomap_2024}. QPCA speedup on quantum data is highlighted in~\cite{covariance_qPCA}. However, the majority of the quantum algorithms for dimensionality reduction still require further algorithmic advancements and fault-tolerant hardware to achieve practical relevance. 

Many other learning tasks may benefit from quantum computing. This includes supervised learning, regression analysis being one example, for instance via quantum linear regression~\cite{wiebe_quantum_fitting_2012, wang_quantum_regression_2017, zhao_quantum-assisted_gpr_2019, date_adiabatic_regression_2021}. Likewise, unsupervised learning may benefit from approaches such as quantum k-means (QK-means) clustering~\cite{khan_qk-means_2019, arthur_balanced_k_means_2021}
and quantum spectral clustering~\cite{kerenidis_quantum_spectral_2021}. Given the recent rise of generative artificial intelligence (AI) techniques, quantum generative models are receiving renewed interest, including QNNs~\cite{tian2023recent}, quantum Boltzmann machines (QBMs)~\cite{amin_quantum_2018, tuysuz_learning_2024}, and quantum generative adversarial networks (QGANs)~\cite{baglio2024data}. As a final example, natural language processing is an emerging area that has recently progressed through the emergence of quantum natural language processing (QNLP) \cite{guarasci2022quantum, di2022dawn}.

\section{Quantum computing for biomarkers}
\label{sec:qcbio}

Quantum computing is not a silver bullet; it does not enable improvements for every single computational task and thus also not every problem in biomarker research. Therefore, it is important to categorize and narrow down which (important) problems in biomarker research are particularly suited with regard to the application of quantum algorithms. Related work has been conducted with regard to biological sciences \cite{emani2021quantum}, cell-centric therapeutics \cite{basu2023towards}, clinical trials \cite{doga2024can}, digital health \cite{gupta2024quantum}, and health and medicine \cite{floether_2023}.

\begin{figure}[h]%
\centering
\includegraphics[width=\linewidth]{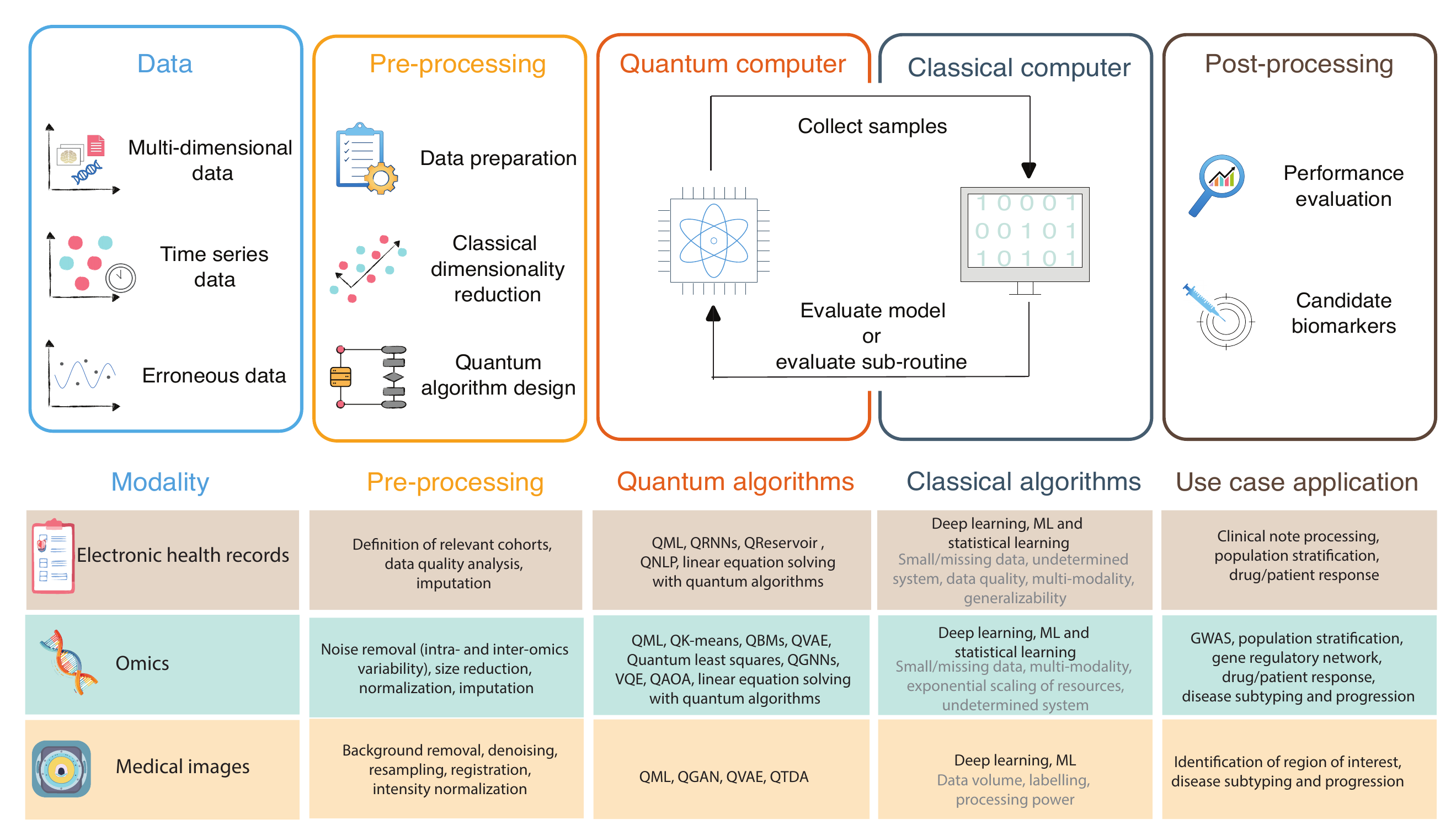}
\caption{Flow of data through a hybrid quantum-classical computational pipeline for different types of data and different healthcare data modalities. The top part of the figure focuses on the categories multi-dimensional, time series, and erroneous data. The bottom part covers the modalities EHRs, omics, and medical images. Examples are given for pre-processing steps, quantum algorithms, classical algorithms and their limitations (in gray), and medical use case applications.}
\label{fig:fig2}
\end{figure}

In this section, we organize areas of opportunity in applying quantum computing to biomarker discovery along the lines of data types while discussing multiple healthcare data modalities for each data type. We chose this categorization instead of one explicitly centered on healthcare data for the following reasons. Quantum algorithm development advances at a fast pace and research in that field often makes explicit assumptions about data types as well as quantum computer accessibility. As such, by focusing on connecting data types to health care modalities, it is our hope that insights provided by this perspective retain relevance in the face of future developments in quantum algorithms. Moreover, this allows insights from the given perspective to have stronger cross-disciplinary relevance. Specifically, in this paper we organize around multi-dimensional data, time series data, and erroneous data. For each of the above data types, their occurrence in and relevance to different healthcare data modalities is analyzed in the context of the applicability of quantum algorithms. Figure ~\ref{fig:fig2} illustrates the flow of data through a hybrid quantum-classical computational pipeline for different types of data, healthcare data modalities, classical approaches and limitations, and pertinent quantum algorithms for selected use cases.

After preparing the data for downstream analysis, classical dimensionality reduction techniques may be applied due to the limited number of available qubits of near-term quantum devices. These may differ between EHRs \cite{sorkhabi2020systematic}, omics data \cite{torres2023omics}, and medical images \cite{math_medical_2024}.

Clearly, the flow illustrates that quantum computing algorithms are only inserted at very specific points of the computational pipeline; most computational steps remain on classical systems. Optimizing the quantum part requires, for example, adapting the algorithm to both the specifics of the device and the characteristics of the data. The classical data should be embedded into a Hilbert space via data encoding techniques \cite{schuld2019quantum}. Depending on the nature of the data and the specific problem at hand, it may be necessary to develop data encoding methods tailored to the application. Additionally, the quantum algorithm should account for the quantum device’s connectivity and its limitation. For instance, error mitigation and suppression protocols~\cite{cai_quantum_2023} should be considered prior to running the algorithm to address potential hardware-induced errors.

\subsection{Multi-dimensional data}
\label{subsec:multi}

Recent technological developments have produced an abundance of data in the domains of healthcare and life sciences \cite{shilo2020axes}. These are often correlated in intricate ways, which may necessitate a (quantum) network model approach~\cite{maniscalco2022quantum} and ML techniques such as (quantum) graph neural networks (QGNNs)~\cite{sagingalieva2023hybrid,ray2023hybrid}.

Furthermore, multi-dimensionality of data can occur due to different modalities and time changes for a set of samples and features. Medical imaging, next-generation sequencing, and other high-throughput instruments produce streams of multi-modal data that are referred to as ``big data''. In the context of biomarker discovery, these massive data volumes acquire orders of magnitude more quantity and complexity when utilized in large-cohort analyses such as genome-wide association studies (GWAS). A key problem, however, arises in the dimensionality of the data. Even big data suffers from having too many features compared with the number of samples, which significantly complicates data science~\cite{prins2017genome}. While there are solutions for increasing the number of samples, these are often still limited for biological datasets. Methods for dimensionality reduction and feature selection therefore become extremely important in handling big data in order to return the dimensionality to where there are many more samples than features. This then blurs the lines between big data and ``small data'', which can either be very limited in the number of samples or the number of features. Small data with few samples is often encountered in population studies of rare diseases and conditions. Even though the underling modality (such as whole-genome sequencing data) is more aptly classified as big data, the intrinsically limited number of samples (for instance in the case of individuals diagnosed with a rare condition) constrains a biomarker discovery investigation to small data. 

In all regimes of data (big, small, and the entire spectrum in between), quantum computing presents us with new opportunities (and a new set of challenges) to address persistent limitations of classical computational approaches. One notable example is overcoming the data and scale inefficiencies of modern classical ML. In the past decade, ML has made incredible advances powered by massively scaling training data, computing budgets, and the sizes of deep neural networks. Recent theoretical results have shown, however, that the efficacy of deep neural networks is tied to the number of samples accessible during the training phase and the effective dimension of the dataset. To illustrate this fundamental relationship, training current neural networks to achieve a robust generalization error on a dataset roughly the size and effective dimension of ImageNet would require a model with around 10–100 billion parameters~\cite{isoperimetry_2021}, rivaling the size of some of the largest large language models trained to-date. Empirical studies by various research groups have confirmed this unfavorable scaling of model size, data size, and computing budget for other data modalities such as natural language~\cite{neural_scaling, compute_optimal}. Recent research in quantum computing points to the possibility of a new class of quantum algorithms that are both data- and model size-efficient and able to generalize to new data with few data samples~\cite{qcnn}. 

While there is evidence for more data-efficient quantum machine learning algorithms, the loading of classical data into quantum computers represents a new set of challenges that exists in both current and near-term quantum hardware. How quantum computers can access data is of central importance in the field of quantum machine learning. In fact, some early quantum machine learning algorithms made assumptions around access to classical data that have been shown to lead to computational advantages over their classical counterparts. But once their data access assumptions were closely inspected and loosened, it was found that many of those computational speedups no longer hold \cite{ewin_dequantized_2021}. Going deeper into the topic of quantum access to data is beyond the scope of this perspective; suffice it to say that for classical big data, data loading may erase computational quantum advantages. Different research efforts are underway, such as QRAMs \cite{phalak2023quantum} on the hardware side and coresets \cite{harrow2020small, yogendran2024big} on the algorithmic side, to address this challenge. For the time being, however, small data is a more natural fit for existing and planned quantum computers.

\subsubsection{Electronic health records (EHRs)}

The increasing adoption of EHRs has revolutionized biomedical research. EHRs provide rapid data availability, eliminating the need for extensive recruitment or data collection. Additionally, they support longitudinal tracking of patient health, facilitating disease progression analysis. Their ability to capture data from diverse patient populations also enhances the detection of subtle health trends.

For the scope of this paper, EHRs are assumed to contain (typically structured) information such as diagnoses, immunization, laboratory measurements, medications, procedures, and vital sign data as well as unstructured doctor notes. They also increasingly integrate social determinants of health (SDOH) and wearable data. Medical images and omics, due to the explosive growth in their collection and relevance, are discussed separately in this paper.

SDOH, encompassing economic stability, education, healthcare access, neighborhood environment, and social context, significantly impact health outcomes \cite{cdc2024social}. While some SDOH data (such as age and location) can be extracted from EHRs, other data (such as social and behavioral determinants) is often missing. Since SDOH are strongly correlated with morbidity and mortality, integrating them with EHRs is crucial. However, the complex interplay between health and social factors presents significant analytical challenges \cite{chen2020social}.

The application of deep learning, machine learning, and AI to EHRs has yielded valuable medical insights, including accurate prediction of future health conditions \cite{ravizza2019predicting, li2020behrt}. AI-based approaches \cite{adamson2023approach, singhal2023opportunities} broaden the understanding of biomarkers, particularly when EHRs are fused with other data sources \cite{mosley2018study}. The construction of such multimodal datasets may itself also benefit from AI techniques \cite{mohsen2022artificial, steyaert2023multimodal}. Therefore, the increasing integration of EHRs in biomedical research suggests that QML holds immense potential for advancing biomarker discovery. In particular, quantum computing could enhance the analysis of EHRs in small-cohort studies, such as clinical trials or rare disease research, by improving predictive accuracy from small data sets.

Furthermore, quantum natural language processing (QNLP) shows promise in representing linguistic information more effectively than classical methods \cite{guarasci2022quantum, di2022dawn}. For example, clinical notes may show that an epilepsy patient has been seizure-free \cite{singhal2023opportunities}, which would otherwise likely have to be indirectly inferred from the absence of medical codes. Therefore, QNLP methods may potentially be applied in analyzing free-text doctor's notes, which offer valuable insights \cite{assale2019revival}.

\subsubsection{Omics}

Now consider omics data. Handling sparse omics data is often a limitation for classic machine learning and statistical techniques in biomarker discovery. Quantum computing may actually provide a solution to some of these challenges.

Genomic, proteomic, and other omics data have revolutionized our understanding of complex diseases. Spurred by technological advances (including next-generation sequencing), omics often includes high-throughput empirical studies, producing raw data that necessitates multi-step computational processing and sophisticated analysis before being amenable to interpretation. In the last decade, omics have become widely used in biomedical research to study disease mechanisms, identification of biomarkers for therapeutic development, and diagnostics in the clinical setting~\cite{tam2019benefits, UffelmannNatReview021}. Although such large datasets can now be generated, biological data suffers from what is known as the ``curse of dimensionality'' \cite{donoho2000high}, which makes it challenging to perform traditional ML, due to over-fitting, and to properly train the model. Mitigation of this is aided by state-of-the-art methods in single-cell sequencing and spatial transcriptomics, which begin to address the issue of dimensionality by interrogating individual cells rather than populations. While the number of patients will not increase with single-cell analysis, having information from individual cells allows for more detailed biomarker identification strategies, especially in diseases with heterogeneous manifestations such as cancer. However, even with single-cell analysis, limited data and too many features continue to be problems, and addressing these requires techniques such as dimensionality reduction, feature selection, and clustering \cite{pudjihartono2022review}.

An example is biomarker discovery for the stratification of patients into different groups based on their susceptibility to a particular disease or their response to a specific drug. Often such studies are limited in the number of available samples, but owing to the large number of features the search space can be very vast. Indeed, collecting omics data from patients is challenging for many reasons, most notably the invasive nature of the medical procedures, their associated costs, and a lack of infrastructure. This is an area where quantum computing has already had demonstrable impact, and will continue to do be increasingly impactful as data availability increases and the hardware and software evolve \cite{nguyen2024biomarker, saggi2024multiomicquantummachinelearning}. 

Furthermore, with genome sequencing becoming increasingly democratized, methods such as GWAS and quantitative trait loci (QTL) mapping are often used to identify genetic biomarkers of disease states or phenotypic variation in populations \cite{albert2015role, tam2019benefits}. However, in order for approaches such as GWAS and QTL mapping to reach statistical significance, let alone train predictive models, huge amounts of data would be required, and this is where quantum computing can provide potential alternatives to classical algorithms \cite{nalkecz2024quantum}. In addition, GWAS studies often require methods such as PCA to identify covariates for biomarker trait regression models~\cite{price2006principal}. In cases where the covariance matrix of interest is of a low rank, QPCA \cite{lloyd_qPCA} and variants thereof \cite{resonant_qPCA, covariance_qPCA} can prove far more efficient than their existing classical counterparts, albeit issues of data loading and quantum hardware fault-tolerance must be overcome first (see discussion in Section \ref{subsec:multi}).

The interpretation of genetic variants in terms of biomarkers is typically limited by inherent correlations between causal and non-causal variants \cite{uffelmann2021emerging, holland2020beyond, slatkin2008linkage}, requiring formal causal inference approaches. The latter are notorious for their prohibitive computational cost \cite{Navascu_CI}, except in cases with restrictive assumptions about the underlying causal structure. Quantum causal inference is an active area of research that could potentially provide more powerful causal discovery algorithms than are currently available classically. So far, quantum speedups in causal inference have been obtained only for datasets whose statistics cannot be described entirely with classical means due to the presence of quantum correlations \cite{chiribella_CI}.

Another prevalent approach to biomarker discovery with omics data is through studies of how transcription of one gene affects the transcription of its neighboring genes, whether it be through statistical approaches such as expression quantitative trait loci (eQTLs) or gaining predictive power from gene expression patterns through machine learning \cite{nica2013expression, khalsan2022survey}. An efficient way to rapidly understand and analyze complex regulatory relationships between genes is through \textit{gene regulatory networks} (GRNs) \cite{chen_identification_2022, zhang_discovering_2022}. GRNs are computationally efficient since they only include pairwise interactions. Recently, a PQC model for a \textit{quantum single-cell GRN} (qscGRN) was proposed to try and allow for higher-order correlations and preserving the computational efficiency of GRNs \cite{roman-vicharra_quantum_2023}. In this work, the authors analyzed single-cell RNA sequencing data from lymphoblastoid cell lines originating from two data sources \cite{Osorio2019,SoRelle2021}. The authors found mixed results with the quantum models, with some correlations in accordance with previous findings, other correlations in contradiction to previous findings, and even new correlations that were not present in the baseline model. This work is an example of how novel quantum computational models and algorithms even, when co-designed with omics problems, can lead to new insights and help bypass the data uploading challenges described in earlier sections.

Moreover, an additional avenue to understand the co-expression of genes is to implement methods to cluster gene-expression datasets. While there are many techniques for clustering gene expression data \cite{Al-Janabee2023-xe}, more traditional methods often make assumptions about the data or make a priori decisions about parameters that are not necessarily sound \cite{Rodriguez2019-wv}. To circumvent this, more complex algorithms such as genetic algorithms or particle swarm optimizers (PSO) have more recently been explored for genomic clustering, with PSO showing promising results with respect to validated accuracies compared with other methods \cite{wang2018particle}. However the disadvantage of this algorithm is that it does not guarantee a convergence to a global minimum \cite{sun2004particle} and is prone to being trapped in local optima \cite{985692}. Classical techniques to overcome this drawback have been discovered, but their implementation can be computationally expensive \cite{10.5555/935867}. A potential solution to mitigate this issue is to use the \textit{quantum-behaved} particle swarm optimization algorithm \cite{sun2004particle, chen2008clustering} and its modified versions \cite{sun2012gene, dabba2020hybridization, Fallahi2022}. The idea is to suppose that each point-particle is now a spin-less one with quantum behavior that follows a corresponding Schr\"odinger equation, which can then be used to perform the particles' position and velocity updates at each iteration. Although the references cited have implemented these quantum-behaved PSO algorithms on classical computers, to our current knowledge, they have not been discussed in the context of being implemented on a quantum computer. It appears possible that furthering the understanding of implementing quantum-behaved PSOs on quantum computers, in addition to exploring other quantum optimization techniques (not necessarily focused on clustering) such as the quantum approximate optimization algorithm (QAOA) and the variational quantum eigensolver (VQE), could lead to quantum advantages.

\subsubsection{Medical images}
\label{subsubsec:mi}

Medical images are a widely used approach for assessing injury, disease, and health. By using techniques such as X-ray, magnetic resonance imaging (MRI), computed tomography (CT), positron emission tomography (PET), and ultrasound, imaging enables physicians and researchers to visualize internal structures and functions of the human body. Qualitative medical imaging involves the visual interpretation of images by trained healthcare professionals to identify and characterize abnormalities, lesions, or anatomical structures by focusing on subjective observations such as size, shape, texture, and density to make diagnostic assessments. This raises issues when it comes to decision-making with images.

Even when looking for known features in medical images, professionals can miss details or find specific images that are difficult to interpret. While algorithms exist for dimensionality reduction and feature selection of images, the amount and diversity of data that needs to be processed can often be a computational challenge and lead to insights being missed. This is where quantum computing could play a significant role. Quantum algorithms \cite{elaraby2022quantum}, particularly quantum machine learning \cite{wei2023quantum}, have shown promise in this application space.

Consider the case where a radiologist may identify the presence of a tumor in an MRI scan based on its appearance and contrast with surrounding tissues. Quantitative medical imaging involves the extraction of numerical or quantitative data from medical images using computer-based algorithms and software tools. Here, subjective measurements of various parameters such as blood flow, density, metabolic activity, and volume are obtained which may allow for more precise and standardized assessments of disease severity, progression, and treatment response. An oncologist may use quantitative imaging techniques that can measure tumor size, shape, and growth rate to assess treatment efficacy and predict patient outcomes. To be able to do this, significant compute power and advanced computational techniques are needed; traditional approaches are often not sufficient. 

Another issue that arises is when specific features are not known, and exploratory analysis is required for diagnosis, which exacerbates the compute problem. For example, although approximately 1.2 \% of the US population has active epilepsy \cite{zack2017national}, only half of these cases have a determined cause. In cases where identification is possible through known markers, MRI is commonly used to detect structural brain abnormalities, such as hippocampal sclerosis, cortical dysplasia, and brain tumors, which may be the underlying cause of epilepsy. In cases where structural abnormalities are not observed, PET can be used to detect areas of hypometabolism that may correspond to epileptogenic zones (EZ). Single-photon emission computed tomography (SPECT) can be used to visualize blood flow changes during a seizure and between seizures, where the comparison of these images can aid in localizing the EZ. Hybrid imaging combines several different imaging techniques, leveraging both structural and metabolic information to improve localization. Extensive analysis is still required, however, to identify more markers that encompass a wider spectrum of epilepsy patients, and also provide more granular detail into the type and severity of the disease, which in turn requires massive amounts of data collection and processing. 

Reconstructing images from lower-dimensional and lower-resolution data is another area of potential quantum enhancement \cite{anuradha2024integrating}. For example, quantum techniques for volume rendering medical imaging data more efficiently have been proposed \cite{yang2019volume}. The challenges in medical imaging go beyond image processing. Medical imaging equipment often comprises devices that require specialized environments and maintenance. For instance, MRI machines need high-powered magnets and must be kept at a temperature of a few degrees kelvin. Lower-powered magnets that do not require such extreme environments, because they benefit from quantum algorithms that place lower demands on image resolution, would be a boon for accessibility and equity.

\subsection{Time series data}

The models used in predicting future values of a time series given a set of past values are fundamentally different to predicting labels or a quantity given a set of features. This is because the former is an extrapolation whereas (in most cases) the latter is an interpolation. Thus, common models (machine learning or otherwise) used in classification or regression problems usually do not perform well for time series prediction and forecasting, and time series data requires other approaches.

For many biomedical applications, analysis of time series data, also known as longitudinal data, presents a range of challenges for classical techniques, including machine learning models. Some of these challenges are algorithm selection, explainability, heterogeneous data, inconsistent labelling and annotations, missing data, and unbalanced data \cite{cascarano2023machine}. In addition, the many correlated variables typically found in biomarker data make it difficult to elucidate causal pathways \cite{snyder2020social}. Quantum computing techniques, particularly QML algorithms such as quantum reservoir methods \cite{mujal2023time}, are actively researched in order to achieve enhanced time series processing \cite{daskin2022walk}.

There already exists a variety of techniques to tackle time series problems. For many examples, one can often get reasonable predictions by using relatively simple models such as auto-regressive moving average models and their many variants \cite{benjamin2003generalized}. For time series data with additional challenges (such as the ones mentioned at the start of this section), the current state-of-the-art is arguably the long short-term memory (LSTM) ML model \cite{van2020review}, which is able to capture some properties, such as seasonality and highly correlated variables, much better than the more traditional models. However, a recent QML technique that has also started to gain attention is quantum reservoir computing (QRC) \cite{mujal2021opportunities}. Reservoir computing is a predecessor to recurrent neural networks (RNNs), where the hidden weights can be treated as a \textit{reservoir} that projects the data into a higher-dimensional subspace, which can then be fed into a simple linear model; quantum analogues exist \cite{takaki2021learning}. The reservoir (similar to the RNN) receives the time series data sequentially, and hence the dynamics of the reservoir are dependent on the previous time step. This reservoir can have many different forms, ranging from classical architectures like the echo state network \cite{lukovsevivcius2012practical} to physical implementations \cite{fernando2003pattern, tanaka2019recent}. The advantage of reservoir computing versus RNNs is that since the majority of the randomly initialized parameters of this reservoir is fixed from the start, the computational cost of training them is much lower. On the other hand, this randomness typically requires reservoirs to have many more parameters to achieve a comparable accuracy. Still, the use of a \textit{quantum} reservoir (for example, by generating a random quantum circuit) can decrease the number of parameters required for the model to perform well.

\subsubsection{EHRs}

Time series analysis is particularly relevant to EHRs, given the longitudinal nature of many of the components, for example laboratory values. To move closer to proactive medicine, one needs to be able to accurately predict the next points in a time series in order to achieve early diagnostics and interventions. EHRs suffer from many of the data challenges described earlier. For instance, many of the patients go in and out of healthcare systems and, as a result, approaches such as hidden Markov models (HMMs) are applied \cite{komariah2019health}. Although these models are good candidates to handle noisy and highly variable signals, they tend to fall short due to several reasons, including the high memory requirements to implement these models at scale and trying to take into account all the important variables. Using QRC or even a quantum version of HMMs could relieve this bottleneck around high-memory requirements if the data can be encoded in an efficient way.

Furthermore, wearables represent another modality that is growing rapidly and which is starting to be integrated with EHRs \cite{dinh2019wearable}. These include chemical sensors that collect signals with electrochemical and optical methods \cite{sempionatto2022wearable} as well as fitness trackers and other devices which monitor digital biomarkers \cite{kourtis2019digital}. They may also include quantum sensors that exploit quantum mechanical properties in order to achieve superior sensitivities, such as nitrogen–vacancy-centre-based magnetometry at the level of single neurons \cite{aslam2023quantum}. By their nature, these wearables typically generate time series data and could thus benefit from such quantum (machine learning) methods.

\subsubsection{Omics}

Omics time series data is particularly valuable for biomarker discovery, as it allows one to observe temporal patterns and gene-gene/protein-protein interactions. Examples includes disease progression/sub-typing~\cite{Hershberger2023, Mihajlovic2024} and patient response to treatments~\cite{Parry2023, 10.1158/1538-7445.AM2023-3874, NAEEM20234623, Burr2024}.

Non-invasive approaches to use liquid biopsies allow one to capture a mixture of the omics profile of cancer cells in a patient via cell-free DNA (cfDNA). Recently, studies have demonstrated the application of cfDNA to identify drug resistance mechanisms~\cite{Khan2018}. Moreover, in the last few years several studies have shown that cfDNA fragmentomics’ characteristics differ in normal and diseased individuals without the need to distinguish the source of the cfDNA fragments, which makes it a promising novel biomarker in particular for early cancer detection~\cite{DNAfragmentomes, Rolfo2023}. Indeed, ML algorithms have been applied to such tasks with promising results to identify late-stage cancer~\cite{DNAfragmentomes} but still fail to detect early signs in the development of the disease. Another recent application of omics time series is in the case of epilepsy post-surgical seizure recurrence~\cite{Hershberger2023} where omics data were used to subtype patients based on their treatment responses.

In general, classical ML often cannot be easily applied to omics-based time series data due to issues such as small sample numbers or missing data. Therefore, data imputation is required, and current imputation techniques often do not incorporate time-dependence, thereby leaving out critical information during imputation. Recent work has demonstrated that graph networks that model topology across samples can be hugely beneficial in such cases \cite{shi2021netoif}. An example is the use of dynamic Bayesian networks for analysis of time series microbiome data \cite{ruiz2021dynamic}, which has been demonstrated to be a critical biomarker for many diseases states \cite{hou2022microbiota, buytaers2024potential}. Still, these topological methods quickly become computationally infeasible with increasingly high-dimensional and complicated data sets, which is where QML and quantum topological data analysis (QTDA)~\cite{QCTDAArxiv} approaches may be able to help, providing an orthogonal point of view and insights. Indeed, they could be used to differentiate with high accuracy between healthy individuals and cancer patients in early stages of diseases, leading to more targeted and effective treatments. 

\subsubsection{Medical images}

Medical image time series data occurs across a continuum of modalities, ranging from a set of snapshots separated in time to movie-like videos. Acquiring image time series data for biomarkers involves capturing medical images over time to monitor changes in biological or physiological states. Examples of time-lapse series include a) MRI or CT scans taken over weeks or months showing the size and shape of a tumor changing in response to treatment and acting as a biomarker for cancer therapy efficacy, b) retinal imaging that tracks progression of diabetic retinopathy via blood vessel leakage, aneurysms, and hemorrhages, and c) PET scans of the brain showing changes in glucose metabolism or amyloid plaque accumulation for neurodegenerative diseases such as Alzheimer's disease. Examples of movie-like imaging data include four-dimensional ultrasound scans (the fourth dimension being time) for fetal development and function and functional MRI to observe changes in brain activity in response to stimuli. 

QRC techniques are relevant techniques for application to these problems. However, the quantum computers would have to be larger than the current generations or the initial image data would have to be downscaled with classical pre-processing techniques, such as passing the image through a residual network \cite{He2015-pr} before sending it through the QRC model. Further investigation is required to understand whether the loss of information during the classical pre-processing would negate the potential advantages of using QRC and related quantum methods instead of classical techniques.

\subsection{Erroneous data}

Noise and errors often appear in the context of quantum computing due to the fragility of the quantum systems and their inherent tendency to decohere. A variety of approaches is being researched in order to handle, or even leverage, such noise \cite{du2021quantum, resch2021benchmarking, peters2021machine, kiwit2024benchmarking, monzani2024leveraging}
 \cite{monzani2024leveraging} and progress towards quantum error correction \cite{sivak2023real}, arguably the holy grail of quantum computing. The focus of this discussion, however, is on handling noisy datasets where the labels are incorrect or missing, which represents a major challenge for classical algorithms \cite{karimi2020deep, algan2021image, song2022learning}. Such misannotations are also of concern when data is synthetically generated \cite{liang2022advances}, an area that is also being explored with QGANs and related methods \cite{khatun2024quantum}.

\subsubsection{EHRs} 

EHR data has been notoriously difficult to analyze due to factors such as errors, lack of interoperability, record gaps, and unusually large or small data sizes. While leveraging such data may also allow better generalization behavior through classical models \cite{ravizza2019predicting}
QML has shown promise in dealing with such difficult datasets even more effectively. For example, the usual presence of quantum noise may actually lead to a greater degree of global robustness and thus fairer models \cite{guan2022verifying}. Moreover, quantum transfer learning may be beneficial when the (real-world) dataset of interest is particularly low- or high-dimensional \cite{otgonbaatar2023quantum}. Quantum support vector machines have been applied to EHRs and have, for small datasets, been shown to be competitive with classical approaches in classifying ischemic heart disease \cite{maheshwari2023quantum} and predicting the persistence of rheumatoid arthritis patients~\cite{krunic2022quantum}. Small datasets, for instance for rare-disease or other very restricted cohorts, may in fact lead to some of the earliest quantum advantages, given that quantum generative and other QML models are particularly suitable for such a setting~\cite{caro2022generalization, hibat2024framework}. Finally, the process of imputation itself may be enhanced, leveraging quantum determinantal point processes~\cite{kazdaghli2024improved}.

\subsubsection{Omics}
Erroneous omics data, often caused by sequencing errors, incomplete data, or misinterpretations, can significantly hinder accurate analysis. These inaccuracies can lead to misdiagnosis, ineffective treatments and loss of trust in omics data. Quantum algorithms can enhance pattern matching/discovery and error correction in omics sequences, making it possible to identify and correct errors~\cite{Charkiewicz2014}. A possible approach could be the use of QTDA which helps characterize the shape and structure of the data in a high-dimensional space (involving Betti number and harmonic representations), making it more effective than traditional methods with regard to identifying and correcting errors in complex omics data. Such techniques not only enhance the quality of the omics data but also accelerate the discovery of biomarkers.

\subsubsection{Medical images}

Erroneous data in medical imaging can lead to misdiagnoses, poor treatment outcomes, and overall inefficiencies in healthcare. Examples include noise, artifacts and image reconstruction issues. Medical images can often display distortions caused by equipment limitations, environmental factors, and patient movements.

Quantum algorithms that have been designed for noise reduction and denoising \cite{wonga2024quantum, wang2024quantum} may be able to more efficiently and effectively process noisy imaging data and distinguish between meaningful information and noise. Due to the large amount of data generated by medical image technology, the compression, storage, and reconstruction of such data remain significant challenges. In many cases, raw imaging data (such as k-space data from MRI) are deleted shortly after acquisition. Quantum-assisted data compression, coupled with QML approaches, may be able to more effectively compress large medical images without losing critical information \cite{haque2023advanced, deb2024quantum}. Finally, as mentioned in Section \ref{subsubsec:mi} image reconstruction is a complex task; it is further complicated by erroneous data. Reconstruction algorithms must be sufficiently robust in order to be able to handle noise and mistakes \cite{ben2021deep}. Quantum algorithms and QML techniques may be more effective at detecting patterns in medical images leading to increased capabilities in indicating and resolving such erroneous imaging data.
 
\section{Open research challenges}
\label{sec:open}
\begin{figure}[h]%
\centering
\includegraphics[width=\linewidth]{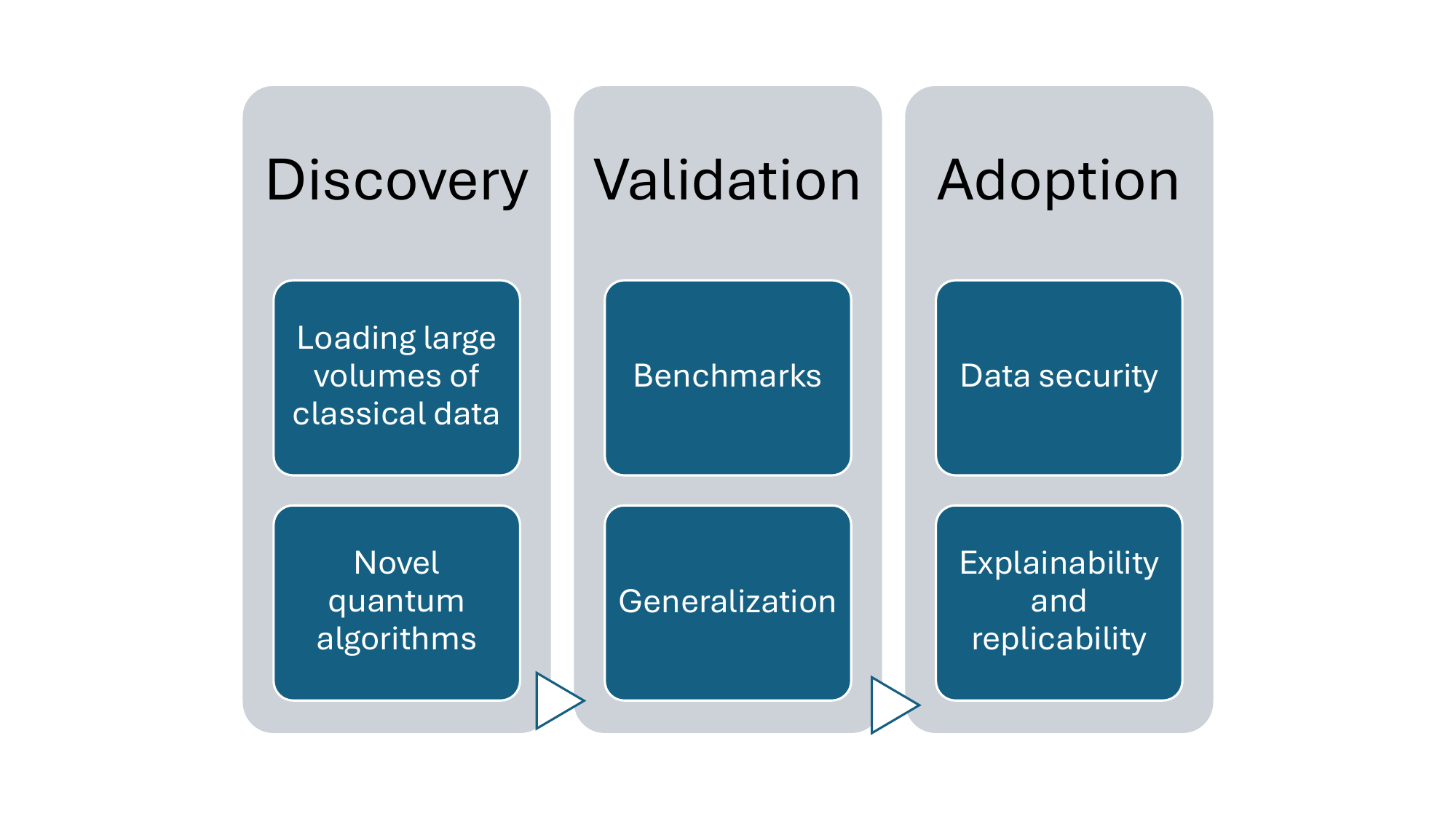}
\caption{Overview of open research challenges in the application of quantum computing in biomarker discovery, validation, and adoption.}
\label{fig:fig3}
\end{figure}

Quantum computers are swiftly moving from ``lab to industry"; nevertheless, a broad range of improvements are still necessary on the road to commercialization and widespread usage. Many of the open problems are cross-industry; better quantum hardware, algorithms, and software will enable more and more use cases. In this section, some of the challenges which are particularly relevant to biomarker discovery, validation, and adoption are outlined; an overview is provided in Figure ~\ref{fig:fig3}.

In the space of biomarker discovery, the loading of data is one key challenge. As described before, biomarker discovery often necessitates extracting insights from large classical datasets. Despite progress with QRAMs, breakthroughs in this area will unlock many more applications. Related to that, quantum algorithms remain an intensely researched area. New ones are being continually discovered; for instance, a study recently claimed an exponential advantage in pathfinding for graphs \cite{li2024exponential}. Furthermore, quantum federated learning represents another novel algorithmic area \cite{li2021quantum, chehimi2022quantum, huang2022quantum, bhatia2023federated}. The idea is to enhance federated learning models where different entities train a model while keeping their data decentralized. Given the growing importance federated learning has for healthcare and medicine, from allowing multi-institutional collaborations without patient data sharing \cite{sheller2020federated} to enabling digital health \cite{rieke2020future} to enhancing the internet of medical things \cite{prasad2022federated}, this is a clear innovation opportunity.

In the area of biomarker validation, a critical aspect concerns benchmarks. The challenge with demonstrating value from quantum computing is that the (classical) goal posts are often not clearly defined and may shift (particularly in response to, or even inspired by, quantum algorithm advances). This makes it challenging to justify the efforts required to develop and use quantum algorithm-based tools. Connected with that, generalization is a key issue due to individual biomarker variability resulting from genetic and lifestyle factors \cite{enroth2014strong}. As a result, more granular (quantum) models will have to be developed, which might again require more (classical) data and novel quantum algorithms.

Finally, data security represents one essential area with regard to biomarker adoption. The high sensitivity of much medical data requires strict data processing standards and is not yet compatible with many quantum computing architectures that demand cloud transfer of data across the world. Similarly, clinical adoption requires a significant amount of trust to be built, which in turn necessitates explainable models and easily replicable results. This challenge, which is also closely linked to similar problems with AI adoption, is exacerbated due to the abstract and inherently probabilistic nature of quantum computing and many of the algorithms.

\section{Conclusion}
\label{sec:conc}

In the long term, the hope is that discovery of better biomarkers helps pave the way towards proactive precision medicine. As opposed to lengthy reactive treatments, each individual may have a continually updated health status that is based on personalized biomarkers and indicates whenever interventions of any form could be advisable. To make this possible, biomarkers must not only be accurately connected with such a health status but must also be accessible in a cost-efficient and easy manner. This is very challenging for diseases, including multi-factorial one such as Alzheimer's disease and cancer, and rare conditions for which little data exists. These are areas where quantum computing applications are particularly promising.

Clearly, there are also still discrepancies with regard to the access to advanced forms of computing, such as quantum computing, across the world \cite{boakye2023state}. While quantum computing, in particular, has been rapidly made more widely available, it is likely to remain relatively expensive for some time to come; today, typical costs are thousands of US dollars per QPU hour. Thus, democratization of such transformative technologies is an important issue that also needs to be considered in order to realize their full impact across the world \cite{seskir2023democratization}.

\section{Acknowledgments}
Sandia National Laboratories is a multimission laboratory managed and operated by National Technology \& Engineering Solutions of Sandia, LLC, a wholly owned subsidiary of Honeywell International Inc., for the U.S. Department of Energy’s National Nuclear Security Administration under contract DE-NA0003525. This paper describes objective technical results and analysis. Any subjective views or opinions that might be expressed in the paper do not necessarily represent the views of the U.S. Department of Energy or the United States Government.

\bibliography{sn-bibliography}

\end{document}